\documentclass{nature}
\usepackage{authblk}  
\usepackage[english]{babel}
\usepackage{bbold}
\usepackage{bbm}
\usepackage{latexsym,amsmath,verbatim}
\usepackage{color}
\usepackage{rotating}
\usepackage{multirow}
\usepackage{color}
\usepackage{amssymb,amsfonts,amsmath}
\usepackage{lineno}

\usepackage{xcolor}
\definecolor{RoyalBlue}{HTML}{4169e1}
\definecolor{ForestGreen}{HTML}{228b22}

\usepackage[colorlinks = true,
        linkcolor = blue,
        urlcolor  = blue,
        citecolor = blue,
        anchorcolor = blue]{hyperref}

\usepackage{subfigure}
\usepackage{comment}

\usepackage{todonotes}

\providecommand{\keywords}[1]
{
  \small	
  \textbf{\textit{Keywords---}} #1
}

\title{Environmental conditions and human activity nexus.\\
The case of Northern Italy during COVID-19 lockdown}

\author
{Sebastian Raimondo$^{a,c}$, Barbara Benigni$^{b,c}$ and Manlio De Domenico$^{c}$}

\begin{document}

\maketitle

\begin{affiliations}
	\item{Department of Mathematics, University of Trento, Via Sommarive, 14, 38123 Povo (TN), Italy}
	\item{Department of Information Engineering and Computer Science, University of Trento, Via Sommarive, 9, 38123 Povo (TN), Italy}
	\item{CoMuNe Lab, Fondazione Bruno Kessler,
		Via Sommarive 18, 38123 Povo (TN), Italy}
	
\end{affiliations}

\noindent $\ast$ Corresponding authors: mdedomenico@fbk.eu, sraimondo@fbk.eu
\date{}

\baselineskip24pt

\begin{abstract}
During COVID-19, draconian countermeasures forbidding non-essential human activities have been adopted worldwide, providing an unprecedented setup for testing sustainability policies. We unravel causal relationships among 16 environmental conditions and human activity variables and argue that, despite a measurable decrease in NO$_2$ concentration due to human activities, locking down a region is insufficient to significantly reduce emissions.  Policy strategies more effective than lockdowns must be considered for pollution control and climate change mitigation.
\end{abstract}

\bigskip


\bigskip

\section*{Introduction}

\begin{figure*}[t]
	\centering
	\includegraphics[width=\columnwidth]{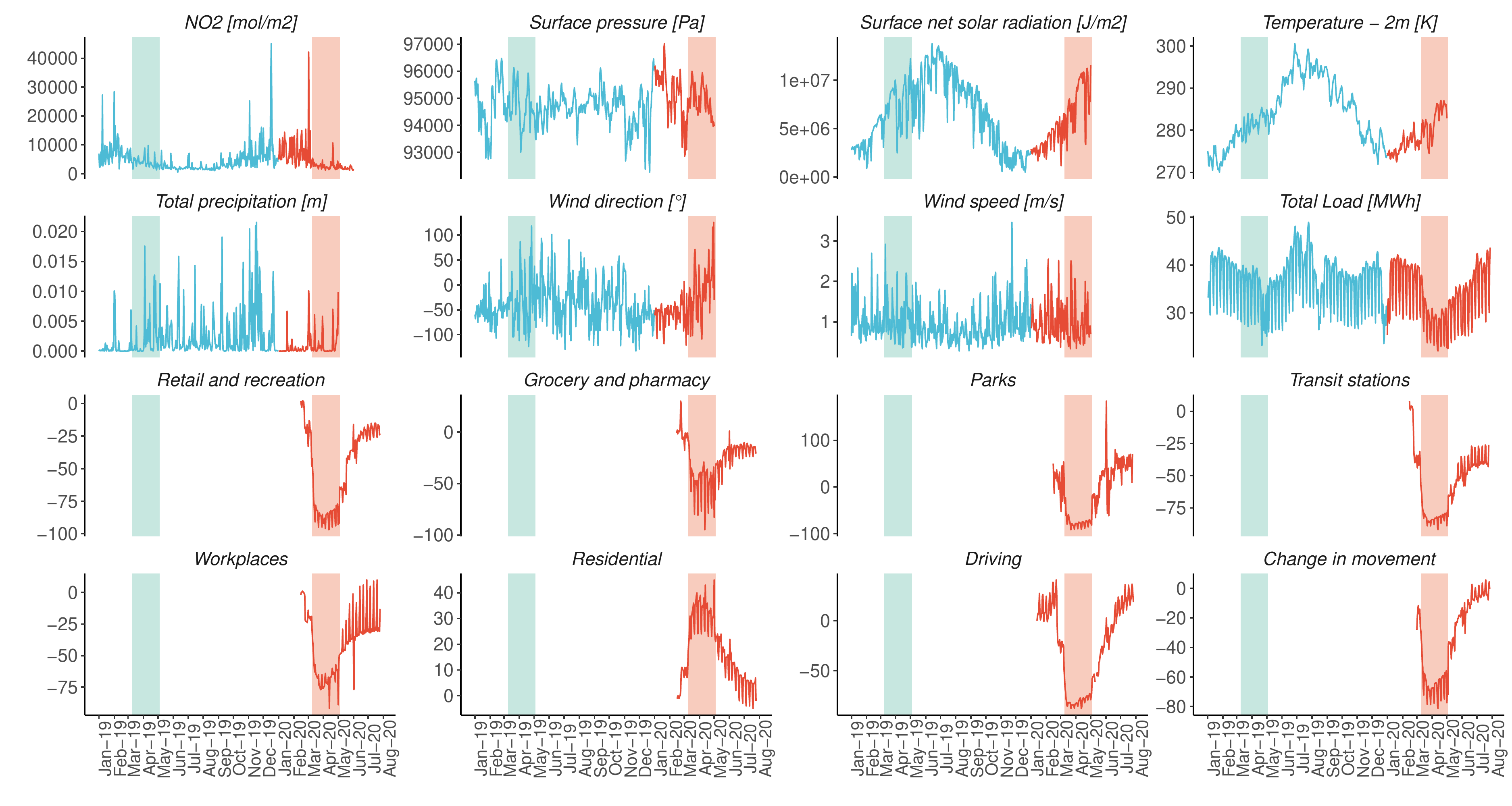}
	\caption{\label{fig:variables}\small{\textbf{ Observables used as a proxy for measuring variations in environmental conditions and human activities during the Italian lockdown.} Panels show the time series for each of the 16 variables considered in this study; blue and red curves corresponds to the times courses for year 2019 and 2020, respectively. The light-red band corresponds to the lockdown period in 2020, while the light-blue band corresponds to the same period in 2019. The public mobility data are available only for the year 2020.}}
\end{figure*}

During the first months of 2020, Italy has been one of the country mostly affected by COVID-19~\cite{zhang2020changes,wu2020nowcasting,salje2020estimating,gonzalez2020introductions,prather2020reducing,ruktanonchai2020assessing,okell2020have,metcalf2020mathematical}.
To prevent the spread of the SARS-CoV-2 virus, Italy adopted significant non-pharmaceutical interventions~\cite{chinazzi2020effect,dehning2020inferring,davies2020effects,haushofer2020interventions}, including locking down the entire country from 9$^{th}$ of March to 4$^{th}$ of May.
The forced closure of school, public facilities and places of employment drastically reduced the vehicle traffic and the industrial activities, with the most relevant effects in the Lombardia region, in the North of Italy, which is also the Italian region most plagued by the virus.
On the one hand, 
this unprecedented situation triggered a cascade of public health, social, behavioral and economic challenges that will require years to recovery. On the other hand, from a scientific perspective, one can investigate the effects of such dramatic regional and sub-regional interventions on the environment. In practice, the spread of COVID-19 allows one to better understand the potential effects of policies that can be adopted to mitigate the climate crisis. To this aim, we consider this situation as a global experiment, giving us a unique opportunity to investigate the complex interactions between human activities and environment with a particular focus on air quality, using the North of Italy as a case study where the availability of heterogeneous data sources allows one to perform a more integrative sustainability analysis. The interest in such a relationship has exploded worldwide during the lockdown period and many recent studies focus on the reduction of atmospheric pollutants due to mobility restrictions in different countries~\cite{tobias2020changes, kerimray2020assessing, bao2020does, Venter}. However, some work highlights that reductions in human mobility and in industrial activity would not be sufficient to reduce air pollution, especially when meteorology is unfavorable~\cite{wang2020severe}.
In fact, it must be noticed that air quality does not only depend on emissions (and consequently concentrations) of pollutants but it is strongly influenced by meteorology, which plays a crucial role in the Po Valley, where Lombardia region is located. Moreover, the effect of the topography, with the Alps to the north, contributes in making this region one of the most polluted areas around Europe.
Our purpose here is to unravel the complex interactions between environmental conditions and human activity in the Lombardia Region, during the Italian lockdown. In particular, we aim at evaluating the impact of relaxing non-essential human activities -- i.e., those activities not directly related to the supply of goods and commodities -- on tropospheric NO$_2$ concentrations during the lockdown.
To this purpose, we compared representative data for environmental conditions -- i.e  air pollution and meteorological conditions data -- and human activity -- i.e. mobility and energy load data -- over a survey period of four months -- February, March, April and May -- both in 2019 and in 2020, for a total of 16 variables (Fig.~\ref{fig:variables}). 
	
By taking into account all these variables we reduce the possible disturbances due to latent confounders. Afterwards, we evaluated the differences in the dynamics of such variables between the period related to a situation with extremely reduced activity and baseline periods. 
Since the lockdown imposed social distancing and, consequently, drastically reduced human mobility, we referred to an indicator of air quality which is sensitive to mobility changes, i.e. the \textit{nitrogen dioxide}~(NO$_2$) concentrations, which proves to highly depend on emissions from transportation means and industrial activity~\cite{lombardia2002rapporto}. 
Concerning the meteorological conditions, we considered the daily average of 6 variables -- i.e surface pressure, surface net solar radiation, temperature, total precipitation, wind direction and wind speed. 
We used the data made publicly available by Google~\cite{google}, Apple~\cite{apple} and Facebook~\cite{facebook} during the survey period to investigate human mobility changes.
In Lombardia region, energy consumption is mainly attributable to industry, consequently we have considered the daily average electricity system’s total load, as a proxy of industrial activity (see Methods for details).
 
From a methodological point of view, we evaluated the significance and the concomitance of variations in human activities and  
in environmental conditions -- focusing on NO$_2$ concentrations -- in the survey period 2019-2020, using statistical and causal analysis. Through statistical tests and effect size measures~\cite{mcgraw1992common, coe2002s} we were able to assess the relevant variations in the time series of the considered variables.
To investigate the complex nexus between the 16 variables we leveraged on the partial correlation coefficient (PCC)~\cite{brown2005partial} and Granger causality (GC)~\cite{granger1969investigating}. 	
Finally, we assessed the robustness of the results found with statistical tests, by taking inspiration from a relatively recent Bayesian technique, based on a state-space model, used to infer the causal impact of advertising campaign on the market sales~\cite{brodersen2015inferring}. By specifying which period in the data should be used for training the state-space model (pre-intervention period) and which period for computing a counterfactual prediction, this technique assesses the impact of the attributable intervention~\cite{brodersen2015inferring}.

\section*{Results}

Figure~\ref{fig:variables} provides an overview of the time course of each observable used during the survey period.
The main result is that the average NO$_2$ concentration during the lockdown period in 2020 is smaller than the one during the same period in 2019 (see \textit{Methods} for the exact start and end dates of the periods used in the statistical tests) . 
It is worth noticing that, in general, at the end of the winter season the NO$_2$ concentration is expected to decrease due to the reduction of the domestic, civil and industrial heating concurrently with the rising temperatures. In 2020 the reduction appears to be more abrupt with the beginning of the lockdown and the concentration levels remain lower for the rest of the period. The variability of the concentration during the lockdown is also lower than the one in the same period in 2019 (Fligner test p-value: $=3.514\cdot10^{-7}$) this may be due to the reduction in the fast-changing pressure variables, such as transportation.
Results of t-tests show that even though the NO$_2$ concentration in the pre-lockdown periods (in 2019 and in 2020) are statistically equivalent (equivalence hypothesis cannot be rejected with p-value $= 0.53$), the reduction observed during the 2020 lockdown with respect to 2019 is statistically significant (p-value $= 7.39\cdot 10^{-4}$). These results are confirmed by a non-parametric surrogate test analysis, leading to a pre-lockdown p-value $= 0.5237$ and to a lockdown p-value $= 2.8\cdot 10^{-4}$. 
To evaluate the magnitude of the observed differences we computed two effect-size measures: the Cliff-$\delta$ and the C.L.E.S. (Common Language Effect Size). For the pre-lockdown periods,
these measures indicate a very low effect size ($\delta \simeq 0$ and $CLES \simeq 0.5$), whereas it is very high for the post-lockdown periods ($\delta \simeq 0.7$ and $CLES \simeq 0.806$). This last result can be interpreted as the probability that an observation from the 2020 lockdown period returns a lower value of NO$_2$ concentration w.r.t. the same period in 2019 is more than 80\%.
Statistical tests on meteorological variables exclude the possibility that the average meteorological conditions in the lockdown period of the two years 2019-2020 are different. We can therefore hypothesize that the environmental conditions were similar and that the variation in the NO$_2$ concentration was driven by the change in human activities. This surmise is further investigated through causal impact analysis.
Finally, the total load in the lockdown period in the year 2020 is significantly smaller than that recorded in the year 2019 (p-value $<10 ^ {- 6}$), while there is no significant difference in the pre-lockdown periods.
It is not possible to test the differences in mobility trends due to lack of publicly available data for the year 2019. 

\begin{figure*}[!t]
\centering
\includegraphics[width=0.98\columnwidth]{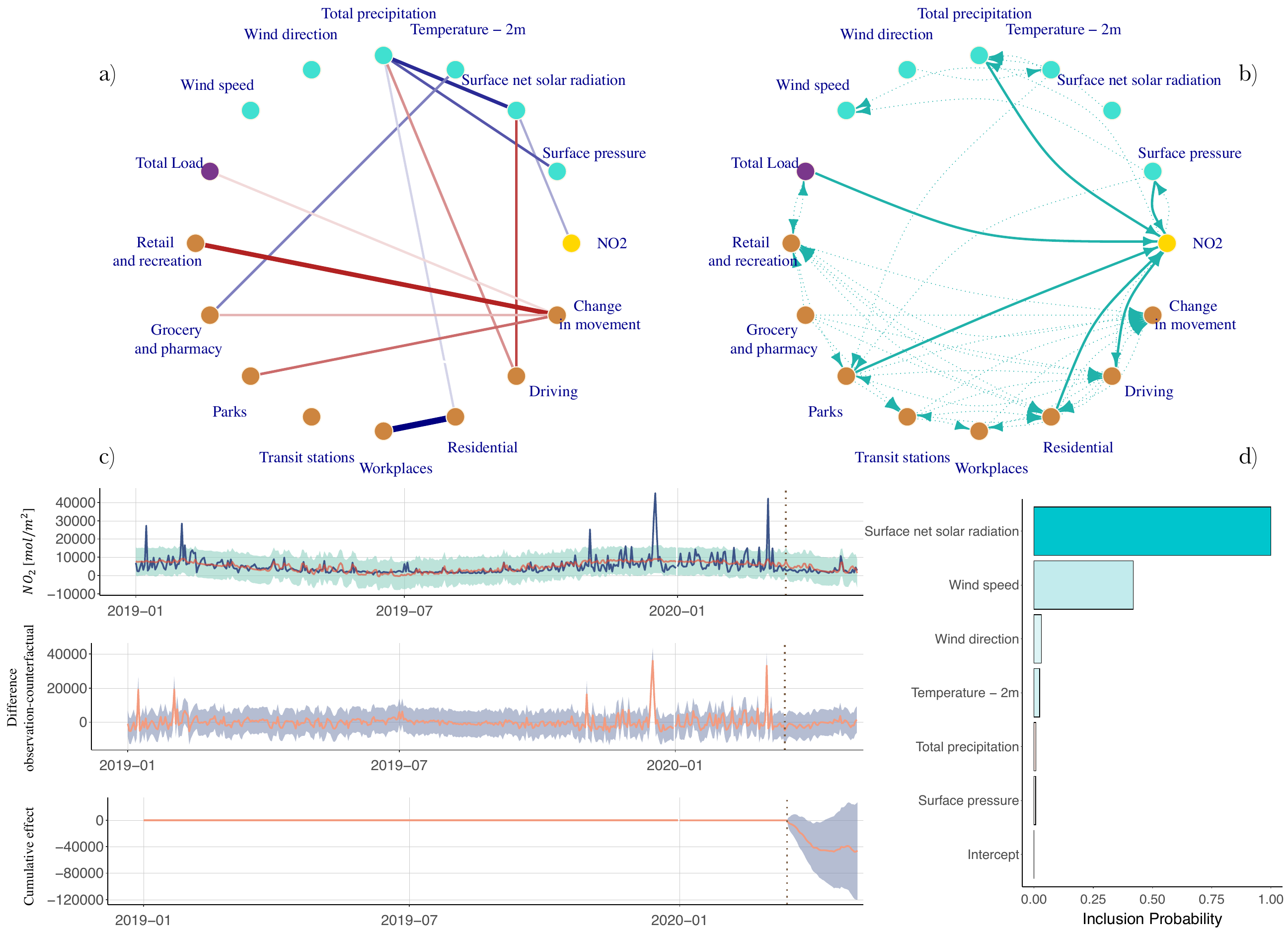}
\caption{\label{fig:results}
\footnotesize{{\textbf{Causal analysis for the time course of the 16 observables.} \textbf{a)} Partial correlation network: each node corresponds to a variable, the color encodes the type of variable (meteorology, energy, mobility, NO$_2$); blue edges represent the negative partial correlation, while red edges represent positive partial correlation; the thickness of the edges is proportional to their partial correlation value. \textbf{b)} Granger causality network: the arrows are oriented in the causal direction; the variable which have a causal impact on NO$_2$ are better highlighted with solid edges. \textbf{c)} Bayesian state-space model: The top panel shows the data and a counterfactual prediction for the lockdown period. The middle plot shows the difference between observed data and counterfactual predictions. The bottom plot is the cumulative effect of the lockdown. \textbf{d)} Probability of inclusion for the regressors; light-blue bars represents negative coefficients, while red bars represents positive coefficients. Note that only the meteorological regressors can be used for the counterfactual prediction, since they are the only variables not influenced by the lockdown intervention.}}}
\end{figure*}

The results of the causal analysis are reported in Fig.~\ref{fig:results}. The partial correlation network in \mbox{Fig.~\ref{fig:results} a)} reveals the stronger relations between all human activity variables, which vary synchronously, driven by the lockdown interventions. In particular the variable ``change in movement" is the most connected. The negative relation between ``residential" and ``work-places" is well captured by the method, likewise for the ``total precipitation", ``surface radiation" and ``temperature". The NO$_2$ variable is related just to the solar radiation, which may be interpreted as the influence of the seasonal effects mentioned in the introduction. 
Figure~\ref{fig:results} b) shows the Granger causality network, which is directed and appears more dense. This network points out the influence of the human activities on the NO$_2$ concentration and also the possible influence of meteorological conditions, that could cause variation on air pollution, such as precipitation. Meaningful meteorological relations can be found in this network, whereas the human activities constitute a dense separated cluster. Results from step-wise linear regression analysis confirm the relation between NO$_2$ concentration, ``total precipitation" and ``parks", and points out also the relation with ``change in movement" and ``grocery and farmacy".

The causal analysis is complemented by the Bayesian state-space model, which includes the meteorological variables as regressors. The results are shown in Fig.~\ref{fig:results} c-d). The plot at the top of Fig.~\ref{fig:results} c) shows the NO$_2$ concentration data and a counterfactual prediction for the lockdown period.
The NO$_2$ concentration during the lockdown period had an average value of $2.76\cdot10^{3}$ mol/m$^2$, whereas in the absence of an intervention, we would have expected an average response of $3.71\cdot10^{3}$ mol/m$^2$. The difference between NO$_2$ concentration data and the corresponding counterfactual prediction (middle plot in Fig.~\ref{fig:results} c) ) yields an estimate of the causal effect of the lockdown. This effect is $-0.96\cdot10^{3}$ mol/m$^2$ with a $95\%$ interval of \mbox{$[-2.55\cdot10^{3}, 0.57\cdot10^{3}]$}.  In relative terms, the response variable showed a decrease of $-26\%$ with a $95\%$ interval of $[-69\%, +15\%]$. 
To obtain the cumulative impact of the lockdown, the concentration data are summed up (bottom plot in Fig.~\ref{fig:results} c) ) obtaining a cumulative concentration of $132.37\cdot10^{3}$ mol/m$^2$ which would have been equal to $178.31\cdot10^{3}$ mol/m$^2$ in absence of the lockdown. 
Even though the results seems to reveal a clear causal impact of the lockdown on the NO$_2$ concentration, the computed probability of obtaining this effect by chance is $p = 0.114$. This suggests that the effect might be due to chance, or to other uncontrollable issues related to the data; for example, because the lockdown period includes data where the effect of the lockdown has already worn off, or because the regressors are not well correlated with the NO$_2$. 

\section*{Discussion}

In this work, we investigated the relationship between human activity and environmental conditions by leveraging the extraordinary event of the Italian lockdown due to COVID-19. To this aim we have fused heterogeneous data sources -- including human mobility, total energy load, meteorological conditions and NO$_2$ concentrations -- in four different periods over 2019 and 2020. We have found evidence that, concomitantly with the reduction in both the mobility and the energy demand, the NO2 average concentration significantly decreases in 2020 lockdown w.r.t the same period in the previous year. The lower variance of NO$_2$ during the 2020 lockdown is even more visible than the one in average concentration. This is attributable to the complex nature of the human-environmental system where internal factors may act as filter on the rapid variations of the NO$_2$. Shedding light on this factor would unveil new possible features of the system under consideration that could be useful for the development of more reliable, data-informed, local climate-change and pollution control policies. 

Given the complexity of the system, where drivers of different nature contribute in affecting the air quality, we determined and quantified the \textit{causal} roles of these drivers on the system itself. Overall, the analysis detected the sign of a causal relation between the relaxation of a broad spectrum of human activities and air pollution abatement  during the lockdown in Northern Italy. However, the statistical significance of the results is questionable, and can be improved with additional data about human mobility and considering additional variables to reduce the noise of latent confounders. Nevertheless, the causal impact may still remain nebulous because of the inherent behaviour of the system.

In the light of these findings, we consider our approach to be indicative, but not definitive, for investigating the nexus between environmental conditions and human activity during COVID-19 lockdown in Italy. Further developments could complement our analysis with mobility data of 2019 and could include more detailed data such as stratified transport (i.e. heavy and light transport). In fact, it is known that motorway traffic has decreased considerably during the lockdown, but the reduction in the circulation of heavy transport has not been so similarly comparable, due to the delivery of essential commodities. For example, in March even though the highway total vehicles dropped by $63\%$, the heavy vehicles decreased by just $27\%$: the latter pollutes four times more than light transport, according to the Italian National Autonomous Roads Corporation (ANAS). 

Here, we proved the power of coupling data analysis with causal inference, to study the nexus between environmental conditions human activity from a systemic perspective. Moreover, we shed some light on the causal relations between human activities and NO$_2$ concentrations, highlighting that a lockdown may be not enough in changing pollutants concentrations and consequently in being regarded as a desirable or even a potential strategy for climate change mitigation and sustainability.   

While the continuity in the essentials supply chain proves the efficiency of the region in providing indispensable services during emergencies such as the COVID-19 pandemic, it also testifies that the backbone of human activities has never really stopped. Consequently, it is plausible to hypothesize that such a backbone might play a more significant role than non-essential one, providing serious challenges to policy and decision making to build a sustainable society in response to climate change.

We firmly believe that a paradigm shift towards a complex systems view is necessary for the optimal management of resources. Such an approach would be of tremendous help for decision making processes allowing for more informed and integrated choices, especially with a view to the development of mitigation policies in accordance with climatic and environmental goals (e.g. Sustainable Development Goals of Agenda 2030).




\section*{Methods}
In this work we relied on data of nitrogen dioxide~(NO$_2$ concentrations) from Copernicus Sentinel-5P satellite from the 1$^{st}$ of January 2019 to the 1$^{st}$ of June 2020 (TROPOMI Level 2 Nitrogen Dioxide total column products. Version 01. European Space Agency~\cite{NO2} ). 
In particular, we referred to high-resolution daily concentrations of the tropospheric NO$_2$ 
over Lombardia region.   
Data of meteorological conditions are retrieved from the Copernicus Climate Change Service~\cite{meteo} (ERA5-Land reanalysis dataset) and consist of hourly data of 6 variables -- i.e surface pressure, surface net solar radiation, temperature, total precipitation, wind direction and wind speed -- over the Lombardia region. These data were averaged daily.   
Google mobility data \cite{google} are provided in terms of daily length of stay at different places -- e.g. residence, grocery, parks, etc. -- aggregated at regional level. Apple data~\cite{apple} are provided in terms of variation in the volume of \textit{driving} directions requests while Facebook data \cite{facebook} in terms of positive or negative \textit{change in movement} relative to baseline (February 2020).
As for NO$_2$, we considered the case of the Lombardia region covering collectively the time period from 13$^{th}$ of January to the 27$^{th}$ of July 2020.
All these mobility data are available only for the year 2020. 
For what concern energy load, we considered data of Northen Italy for the period from 1$^{st}$ of January 2019 to the 27$^{th}$ of July 2020 from the Italian transmission system operator \textit{Terna}~\cite{terna}. "Northen Italy" is the smallest available space aggregation which includes Lombardia; since Lombardia has the highest energy demand in this area, Northen Italy data are deemed appropriate for our purposes. 

To determine whether data reflect the regime shift of the lockdown we implemented a shifting point detection technique based on an information-theoretic measure of similarity. Through this data-driven procedure we found the date which splits the survey period in pre-lockdown and lockdown periods. This is used both for statistical tests and for the Bayesian state-space model. In particular, we applied the Jensen-Shannon divergence to shifting subsets of the NO$_2$ time series (of length $N$) for the 2020.

The main result of the process is the date of the tipping point in which the dynamics meets a regime shift.
The date of the ``information-theoretic'' lockdown turns out to be the 14$^{th}$ of March (p-value = $6.12 \cdot{10^{-5}}$), 5 days after the institutional lockdown. This result is consistent with the physical behaviour of the NO$_2$ which has a typical lifetime of few days. Consequently, we grouped the time series in the 4 subsets used for testing: 
\begin{itemize}
\item {Group 1.1 (pre-lockdown) from 1 February 2019 to 14 March 2019}
\item {Group 1.2 (lockdown) from 14 March 2019 to 5 May 2019}
\item {Group 2.1 (pre-lockdown)  from 1 February 2020 to 14 March 2020}
\item {Group 2.2 (lockdown) from 14 March 2020 to 5 May 2020}
\end{itemize}

To rigorously assess the differences in the time series of the considered variables, we tested the differences in mean between each period with t-tests and surrogate data tests. In addition, we evaluated the differences in the variances between the above periods with F-test, Bartlett's test and Fligner test. 

The magnitude of the differences observed in the time series are evaluated through two effect-size measures: the Cliff-$\delta$ and the C.L.E.S. (Common Language Effect Size).
The Cliff-$\delta$ is computed by enumerating the number of occurrences of an observation from one group having a higher response value than an observation from the second group, and the number of occurrences of the reverse:
\begin{equation}
\delta=\frac{\sum_{i=1}^{n_{1}} \sum_{j=1}^{n_{2}} \operatorname{sign}\left(x_{i 1}-x_{j 2}\right)}{n_{1} \times n_{2}}
\end{equation}
where the two time series are of size $n_1$ and $n_2$.
The C.L.E.S. instead is defined as the probability that a randomly selected individual from one group have a higher score on a variable than a randomly selected individual from another group. We computed this measure numerically with a brute-force approach.
Afterwards, to obtain a statistical indication of the possible causal relations between the 16 variables, we build the partial correlation matrix. The partial correlation measures the strength and the direction of the (rank) dependence of two variables from a set of random variables when the influence of the remaining variables is removed. 
We further investigated the causal dependence through the ``predictive causality" measure by C. Granger, and we compared the results with a step-wise linear regression with the AIC selection criteria. It is to be noticed that not all the required constraints for Granger causality are strictly respected (i.e. linearity and stationarity of the observations). For this reason and to assess the robustness of the results found with the statistical tests, we employed a Bayesian modeling technique. This technique uses a state-space model to predict what would have been the system evolution after an ``intervention", if the intervention had never occurred. For this method to work correctly, the covariates themselves must not be affected by the intervention and the relationship between covariates and treated time series, as established during the pre-intervention, must remains stable throughout the post-intervention period. This procedure does not require linearity nor stationarity in the data but, in turn, it cannot find a direct causal nexus between NO$_2$ concentration and the variables affected by the lockdown measures. In our case, the idea is to detect a possible deviation of the NO$_2$ time series from the counterfactual prediction, when the lockdown is the 
``intervention" and the meteorological conditions are the covariates. If the meteorological conditions were the only variables that influence the NO$_2$ concentration, we would expect the data to follow the counterfactual prediction for the post-treatment period. 
If, instead, other variables (the ones related to human activity restrictions) were responsible for the variation in the NO$_2$ concentrations the deviation from the counterfactual would be clear. 
Thanks to this approach we stated the impact of the reduced human mobility and energy consumption due to the lockdown, considered as the attributable intervention on NO$_2$ concentrations.

\bibliographystyle{naturemag}
\begin{small}
\bibliography{biblio}
\end{small}

\section*{Author contributions statement}
M.D.D. and S.R. conceived the experiment; M.D.D and S.R. conducted the experiment; B.B., M.D.D., S.R analysed the results; B.B, M.D.D. and S.R. wrote the manuscript. All authors reviewed the manuscript. 

\section*{Additional information}
\textbf{Competing interests}: The authors declare no competing interests.

\end{document}